\documentclass[a4paper,11pt]{article}
\usepackage{amsmath,amsfonts,a4wide}
\numberwithin{equation}{section}

\DeclareMathOperator{\prob}{{\mathbb P}}

\DeclareMathOperator{\curl}{curl}

\makeatletter
\newcommand{\eqcolon}{\mathrel{\mathord{=}\raise.2\p@\hbox{:}}}
\newcommand{\coloneq}{\mathrel{\raise.2\p@\hbox{:}\mathord{=}}}
\makeatother

\newtheorem{definition}{Definition} 
\newtheorem{lemma}{Lemma} 
\newtheorem{proposition}{Proposition} 
 
\newtheorem{theorem}{Theorem} 
\newtheorem{remark}{Remark} 
  
\newenvironment{proof}[1][Proof]{{\sc #1.} }{\hfill $\Box$}
\newcommand{\RR}{\mathbb{R}^3}
\newcommand{\R}{\mathbb{R}}
\newcommand{\N}{\mathbb{N}}
\newcommand{\Z}{\mathbb{Z}}

\newcommand{\la}{\langle}
\newcommand{\ra}{\rangle}
\newcommand{\X}{\mbox{$\vec{\bf X}$}}
\newcommand{\U}{\mbox{$\vec{\bf u}$}}
\newcommand{\UU}{\mbox{${\bf u}$}}
\newcommand{\bgamma}{\mbox{$\vec{\boldsymbol \gamma}$}}
\newcommand{\bbgamma}{\mbox{${\boldsymbol \gamma}$}}
\newcommand{\vomega}{\mbox{$\vec{\boldsymbol\omega}$}}
\newcommand{\vPsi}{\mbox{$\vec{\boldsymbol \Psi}$}}
\renewcommand{\k}{\mbox{$\vec{\bf k}$}}
\newcommand{\x}{\mbox{$\vec{\bf x}$}}
\newcommand{\y}{\mbox{$\vec{\bf y}$}}
\newcommand{\m}{\mbox{$\vec{\bf m}$}}
\newcommand{\bxi}{\mbox{$\vec{\boldsymbol \xi}$}}
\newcommand{\bbxi}{\mbox{${\boldsymbol \xi}$}}
\begin{document}
\title{\bf On the global evolution of vortex filaments, blobs, and
  small loops in 3D~ideal flows} 
\author{Luigi C. Berselli and Massimiliano Gubinelli\\
Dipartimento di
  Matematica Applicata ``U.Dini'',\\ 
Universit\`a di Pisa, Pisa I-56126 --
  ITALY \\ \\ 
email: berselli@dma.unipi.it, m.gubinelli@dma.unipi.it\\
Tel: +39 050 2219450, Fax: +39 050 2219479}
\date{}\maketitle
\begin{abstract}
We consider a wide class of approximate models of evolution of singular
distributions of vorticity in three dimensional incompressible fluids
and we show that they have global smooth solutions. The proof 
exploits the existence of suitable Hamiltonian functions.\par 
The approximate models we analyze (essentially discrete and continuous
vortex filaments and vortex loops) are related to some problem of
classical physics concerning turbulence and also to the numerical
approximation of flows with very high Reynolds number. Finally, we
interpret our results as a basis to theoretical validation of
numerical methods used in state-of-the-art computations of turbulent
flows.  
{\small \vskip.3cm 
\noindent{\bf 2000 MSC:} 76B47, 76B03,
45K05.
}  
\end{abstract}
\section{Introduction}
In this paper we consider the Euler equations describing the motion of
an ideal (inviscid), incompressible and homogeneous fluid in $\R^n$
($n=2,\,3$):
\begin{equation}\label{euler}
\frac{\partial \U}{\partial t} +(\U\cdot \nabla)\,\U+\nabla
p={\bf0},
\end{equation}
and
\begin{equation}
\nabla\cdot \U=0,\label{div}
\end{equation}
with the associated initial condition $\U|_{t=0}=\U_{0}(\x),$ where
$\U=(\UU^1,\dots,\UU^n)$ is the velocity field, while the scalar $p$ is
the kinematic pressure.\par 
To solve~\eqref{euler} and~\eqref{div}, the introduction of vorticity
field is crucial. This quantity, that measures the ``amount of
rotation'' inside the fluid, is responsible for the big differences
between the 2D and the 3D case, see Wolibner~\cite{Wol33} and
Yudovich~\cite{Yud63}. If $n=2$, taking the curl of
equation~\eqref{euler}, one gets
\begin{equation*}
{\displaystyle\frac{\partial \vomega}{\partial t} +(\U\cdot
  \nabla)\,\vomega={\bf0},}
\end{equation*}
where 
%
%
$\vomega:=\text{curl}\ \U=(0,0,\partial_1\UU^2-\partial_2 \UU^1)$, 
that is a pure transport equation. Then, a bounded initial vorticity
remains bounded for all the positive times. Let us denote by
$\X_{t}(\x)$ the position at time $t$ of the fluid particle that at
time 0 was in $\x$.  We have the following equation (equations for the
characteristics) 
\begin{equation}
\frac{d\,\X_{t}(\x)}{dt}=\U(\X_{t}(\x),t).\label{characteristic}
\end{equation}
The vorticity field is simply transported along the trajectories of
particles $\X_t,$ $t\geq 0$, remaining always perpendicular to the
plane of motion. Beside, the vector $\vomega$ having two component
identically zero behaves essentially as a scalar.\par 
When $n=3$, the situation becomes much more complicated. The vorticity 
associated to $\U$ is a 3D~vector field satisfying the following
equation
\begin{equation*}
\frac{\partial \vomega}{\partial t} +(\U\cdot \nabla)\,\vomega=
(\vomega\cdot \nabla)\,\U.
\end{equation*}
In contrast to the 2D~case, the vorticity $\vomega$ is
no longer conserved and the fundamental
complication arising in three dimensions is the presence on the right
hand side of $(\vomega\cdot \nabla)\,\U,$ the ``vortex stretching
term,'' which is --roughly speaking-- of order of $|\vomega|^2$.\par 
%
Some numerical simulations suggest that solutions of the ideal fluid
equations may develop singularities and that the possible
singularities may be detected by the blow-up of the vorticity field
$\vomega,$ see Beale, Kato, and Majda~\cite{MR85j:35154} and
references therein. Moreover, accurate direct numerical simulations of
3D turbulent fluids show that the regions where $|\vomega(\x)|$ is big
have an elongated ``filament'' shape, see for instance Bell and
Markus~\cite{BM}, Vincent and Meneguzzi~\cite{VM}, and the reviews in
Chorin~\cite{MR95m:76043} and Frisch~\cite{f}. Thus, a consistent part
of the current research activity is devoted to find suitable
mathematical descriptions of these (and other geometric) vortex
structures, see for instance Constantin~\cite{MR95d:76057}. In recent
years also the use of the Lagrangian approach seems promising, to
detect different intrinsic geometric properties of fluid flows, see
Constantin~\cite{aa}. \par   
We briefly recall some recent contributions on the study of these
filaments, motivated also by application to the numerical simulation:
Chorin~\cite{MR95m:76043} proposed a model based on self-avoiding
random walks, using ideas coming from the theory of polymers;
Gallavotti~\cite{g} suggested the introduction of non-smooth curves
(like Brownian motions), to avoid some of the divergences arising in
the equations; Lions~\cite{l} and Lions and Majda~\cite{lm} proved a
mean field result based on an approximate model of ``nearly parallel''
vortex filaments; Flandoli~\cite{franco} introduced a probabilistic 
description and overcame certain divergences in the energy by
considering filaments with a fractal cross-section. Furthermore, by
using this model, authors in~\cite{FG} have defined
certain Gibbs measures, while Bessaih and Flandoli~\cite{BF} proved a
mean field result. Recent overview on vortex computational method can
be found in Cottet and Koumoutsakos~\cite{ck}.\par
On the other hand, relevant advances in applied mathematics have been
obtained in the study of 2D~models for fluids (like the quasi-geostrophic
equations) that replicate the 3D~vorticity behavior \textit{via}
``analytic and geometric analogies,'' see Constantin, Majda, and
Tabak~\cite{MR95i:76107} and C\'ordoba and C\'ordoba~\cite{cc}. This is
obtained with a smoothing or a modification 
(with appropriate scaling properties) of the Biot-Savart law. Finally,
very recently Cordoba \textit{et al.} (see~\cite{cf} and references
therein) studied the breakdown of smooth tubes of vorticity and other
singularities concentrated on small but elongated sets.\par 
The basic goal of this paper is to show global existence of smooth
solutions for models of line vortices, supplementing previous work
in~\cite{bb}. 
In particular, we shall exhibit a Hamiltonian function. 
The main results will be derived by showing that the energy is
conserved and also that the velocity induced by a singular
distribution of vorticity is smooth enough. We shall consider
rectifiable curves, but similar results can be also be proved for
H\"older curves and some rough paths.\par 
We shall also consider the vortex loop approximation (see
Buttke~\cite{MR1248421} and Chorin~\cite{MR95m:76043}) for the
equivalent equations of motion involving the impulse density. Also for
this model we shall prove similar results of existence for large times
of a smooth solutions.\par    
Finally, we discuss the evolution for approximate
equations in which a smooth vortex filament is approximated by a
discretization and we shall connect the existence of solutions with an
appropriate choice or refinement of 
interior points. This latter model is motivated being the simpler
numerical approximation to the line vortex evolution problem.

\vspace{.5cm}
\noindent\textbf{Plan of the paper:} The paper is organized as
follows: In Section~\ref{section_formal} we recall the motivation for
the introduction of models with vorticity concentrated on singular
sets together with classical methods to obtain integrable
equations. Section~\ref{section_energy} introduces a proper
Hamiltonian and gives the main analytical results: conservation of
energy and estimates of the velocity in terms of the energy that
prove global existence of strong solutions. Section~\ref{Buttke_loops}
uses the same technique to deal also with the problem of vortex loops
in magnetization variables. Finally, in Section~\ref{discrete} a
question connected with the resolution needed to simulate the behavior
of a line vortex is put in the framework of the previous results. 
\section{Vortex motion and approximate models}
\label{section_formal}  
To introduce the problem we shortly recall the discussion in Section~2
of~\cite{bb}, where the reader can find full details. Curves that are
always parallel to the vorticity vector are known as \textit{vortex
  lines}, and vortex lines passing through the points of  a closed
curve define a volume called \textit{vortex tube}.\par 
A \textit{vortex filament} is a vortex tube that is immediately
surrounded by irrotational fluid, even if the usage of this expression
is neither uniform nor consistent. It can be also used to design
infinitesimal vortex tubes or also patterns of vorticity that are not
properly formed by vortex lines.\par
A \textit{line vortex} is a singular distribution in which infinite
vorticity is concentrated on a line, such that the circulation around
a closed circuit threaded on by the line is finite. A line vortex is
the result of a limiting process in which a vortex filament (of finite
strength) is contracted to a curve, the strength $\Gamma>0$ being kept
constant.\par 
The laws of vortex motion, first formulated by Helmholtz~\cite{helm},
state that in an ideal fluid vortex lines and tubes move with the
fluid and that the strength of each vortex tube (flux through any
cross-section) remains constant in time.\par  
Hence, we consider an ideal fluid with the vorticity $\vomega$
concentrated on a smooth curve $\bgamma$. Neglecting the transverse
size of the filament, we can write the vorticity field $\vomega$
generated by $\bgamma$ (that is described parametrically as a
continuous function $\bgamma :[0,1] \to \mathbb{R}^3$ such that
$\bgamma(0) = \bgamma(1)$) and, \textit{formally},
\begin{equation}\label{formal}
  \vomega(\x,t)=\Gamma\int_0^1\delta(\x-\bgamma(\xi,t))
  \bgamma_\xi(\xi,t)\,d\xi\qquad\forall\,\x\in\R^3,\
  \forall\,t\in]0,T[, 
\end{equation}
where $\delta(\,.\,)$ is the Dirac's delta function, $\xi$ is the
arc-length, the parameter $t$ represents the time, and $\bgamma_\xi$
is the derivative with respect to the arc-length.\par
A discussion on the precise meaning of the \textit{formal} vorticity
field~\eqref{formal} and its distributional rigorous definition, can
be found, for example, in the appendix of~\cite{bb}.     
\begin{remark}
The above formula can also be written formally as the distribution
\begin{equation*}
  \vomega(\x,t)=\Gamma\int_0^1\delta(\x-\bgamma_\xi)\,d_\xi
  \bgamma,
\end{equation*}
meaning that the integral could be the Riemann-Stieltjes integral or
also that included in a more general setting. In the sequel we shall
not write explicitly the dependence of $\vomega$ and $\U$ on $\bgamma$
since we shall always consider vorticity (and corresponding velocity)
``generated'' by a distribution concentrated on a curve
$\bgamma$. \end{remark} 
It is well-known that the kinetic energy associated to~\eqref{formal}
  (in the sequel we shall denote by $\la \vec{\bf a},\vec{\bf b}\ra$
  the scalar product in $\R^3$ of the vectors $\vec{\bf a}$ and
  $\vec{\bf b}$) 
\begin{equation*}
E(t)=\frac{\Gamma^2}{8\pi}\int_0^1\int_0^1\frac{1}
{\displaystyle|\bgamma(\xi,t)-\bgamma(\eta,t)|}
\la{\bgamma_\xi}(\xi,t),\bgamma_\eta(\eta,t)\ra\,d\xi 
    d\eta,  
\end{equation*}
is infinite for any reasonably smooth curve $\bgamma$. This explain
the reason for the study of approximate models (as the Localized
Induction Approximation (LIA) or the Rosenhead approximation,
see Saffman~\cite{s}) or for the same problem with curves supported by
non-smooth lines (see Gallavotti~\cite{g} and recent results
in~\cite{BGR}). To derive the evolution equation satisfied by a line
vortex the essential tool is the Biot-Savart formula that gives a
representation formula for the velocity, in terms of the vorticity:  
\begin{equation*}
\U(\x)=-\frac{1}{4\pi}
\int_{\R^3}\frac{(\x-\y)}{|\x-\y|^3}\wedge \vomega(\y)\,d\y, 
\end{equation*}
where $\vec{\bf a}\wedge \vec{\bf b}\in\R^3$ is the vector (exterior)
product of the vectors $\vec{\bf a},\,\vec{\bf b} \in \mathbb{R}^3$.
Inserting the expression~\eqref{formal} of the vorticity field in the
Biot-Savart formula, using the equation for the
characteristics~\eqref{characteristic}, and imposing that the curve
$\bgamma$ is transported by the velocity field $\U$, we get the
following equation
\begin{equation*}
\frac{\partial\bgamma}{\partial
    t}(\xi,t)=-\frac{\Gamma}{4\pi}\int_0^1{\frac{\bgamma(\xi,t)-
\bgamma(\eta,t)}{|\bgamma(\xi,t)-
\bgamma(\eta,t)|^3}\wedge
\bgamma_\eta(\eta,t)}\,d\eta.
\end{equation*}
This equation (relation between its solutions and weak solutions to
the 3D~Euler equations can be found -for instance- in the appendix
of~\cite{bb}) has been used for the first time by J.J.~Thomson, who
proposed a cut-off in the integral in order to control an infinite
energy, see~\cite{s}. An approximation based on the Taylor expansion
of the kernel in the singular integral gives the Localized Induction
Approximation (LIA) solved by Hasimoto~\cite{h}, while recent results
in this direction are those by Klein and
Majda~\cite{MR92d:7602,MR92m:76038}. \par 
The de-singularized model studied analytically in~\cite{bb} is that
proposed by Rosenhead~\cite{Rosenhead},
\begin{equation}\label{eq_rosenhead}
  \frac{\partial\bgamma}{\partial t}=-\frac{\Gamma}{4\pi}
\int_0^1{\frac{\bgamma(\xi,t)-\bgamma(\eta,t)}{\big[\big(\bgamma(\xi,t)-
\bgamma(\eta,t)\big)^2+\mu^2\big]^{3/2}}
\wedge\bgamma_\eta(\eta,t)}\,d\eta,
\end{equation}
for some $\mu>0.$ The choice was motivated by the fact
that~\eqref{eq_rosenhead} gives advantages for numerical integration
and, starting by Moore~\cite{Moore}, it has been used in the applied
context  for some calculation related to aircraft trailing vortices.  
\subsection{On the Rosenhead model}
Regarding the Rosenhead model, a first analytical result has been
recently proved in~\cite{bb}. In that reference is shown (as a
particular case) that if the initial curve is smooth and closed then
the initial value problem is well-posed in natural Sobolev spaces. In
the sequel we shall use the customary $W^{k,p}$ and $H^k=W^{k,2}$
spaces, see Adams~\cite{Ada75}. To our knowledge the following is
essentially the first rigorous existence and uniqueness result, known
for this model. 
\begin{theorem}\label{theorem_existence}[See~\cite{bb}]
Let $\bgamma_0(\xi)\in H^1_\#(0,1)$, where $H^1_\#(0,1)$ denotes the
subset of closed curves $\R\mapsto\R^3$ belonging to the Sobolev space
$H^1(0,1)$. Then, there exist a strictly positive $T^*\leq T$ and a
unique curve $\bgamma$ such that
  \begin{enumerate}
  \item[a)] $\bgamma\in W^{1,\infty}(0,T;L^2(0,1))\cap
 L^\infty(0,T;H^1_\#(0,1));$
   \item[b)]
     ${\displaystyle\bgamma(\xi,t)=\bgamma_0(\xi)-
\frac{\Gamma}{4\pi}\int_0^t\int_0^1{\frac{\bgamma(\xi,\tau)-
\bgamma(\eta,\tau)}{\big[\big(\bgamma(\xi,\tau)-   
\bgamma(\eta,\tau)\big)^2+\mu^2\big]^{3/2}}
\wedge\bgamma_\eta(\tau)\,d\eta d\tau,}}$
  \end{enumerate}
that is a strong solution to~\eqref{eq_rosenhead} in the time interval
$[0,T^*).$\par   
In addition the time $T^*>0$ may be estimated from below and it
depends just on the $H^1$-norm of the initial datum.
\end{theorem}
A similar result has been recently proved in a more general stochastic
context in~\cite{BGR} with a different target and weaker hypotheses on
the smoothness of the curve $\bgamma$. Anyway, in both papers the
question of the global solvability left open and some continuation
criteria involving the length of the curve itself were derived.\par
Our main goal is now to show that, for a  class of regularizations
(including the one above)  the above local theorem is in fact global.
\subsection{A class of regularized problems}\label{regularized_problems}
An interesting class of regularized evolution equations can be defined
as follows. Let $\varphi: \R^3 \to \R$ be a scalar function and define
the velocity field $\U$ associated to the vortex filament $\bgamma$ as 
\begin{equation}
  \label{eq:mod-velocity}
\U(\x):= (\curl \varphi) * \vomega = \int_0^1 \nabla
\varphi(\x -\bgamma(\xi)) \wedge \bgamma_\xi(\xi)\,d\xi,  
\end{equation}
where the convolution $*$ has to be intended in the sense of
distributions. Formally, this formula is such that
$\curl\U=\vomega$.\par 
%
%
On the kernel $\varphi$ we impose the following conditions that will
allow the energy to be well-defined.

\vspace{.3cm}
\noindent\textbf{Hypothesis A.} In the sequel we shall assume that:
\begin{enumerate}
\item[-] The function $\varphi$ is even:
  \begin{equation*}\tag{A.1}
    \varphi(-\x)=\varphi(\x);
  \end{equation*}
\item[-] The function $\varphi$ has a real and non-negative Fourier
  transform: 
  \begin{equation*}\tag{A.2}
\widehat \varphi(\k) := \int_{\RR}  \text{e}^{\,i\la \k,\,\x\ra}
\varphi(\x) \,d\x \ge 0\qquad\forall\,\k\in\R^3;
  \end{equation*}
\item[-] The Fourier transform $\widehat{\varphi}(\k)$ is integrable
  over $\R^3$: 
  \begin{equation*}\tag{A.3}
    \int_{\R^3}\widehat{\varphi}(\k)\,d\k<+\infty.
  \end{equation*}
\item[-] The function $\varphi$ is smooth enough, in order that 
\begin{equation*}\tag{A.4}
\int_{\R^3}  (1+|\k|^2)^2 \widehat{\varphi}(\k)\,d\k < \infty; 
  \end{equation*}
\end{enumerate}
These assumptions are satisfied by a wide range of function $\varphi$,
covering also well-known cases with a physical meaningful
interpretation. 
\begin{remark}[An explicit relevant example]
The function 
$$
\varphi_R(\x) = \frac{\Gamma}{(|\x|^2+\mu^2)^{1/2}}
$$ 
for some $\mu > 0$ would be a natural candidate satisfying all the
above assumptions. Assumption (A.1) is trivially satisfied. Regarding
the assumption (A.3), the Fourier Transform may be expressed in terms
of a Bessel function of the second kind $y(z)=\mbox{\rm Bessel}(1,z)$,
defined as solution of the differential equation
$z^2y''+zy'-(z^2+1)y=0$. The condition on the non-negativity of
the Fourier transform can be verified by a direct computation: 
\begin{equation*}
  \begin{aligned}
\widehat{\varphi}_R(\k) & = \Gamma \int_{\R^3} \frac{\mbox{\rm e}^{\,i\la
    \k,\,\x\ra}}{(|\x|^2+\mu^2)^{1/2}}\,d\x \\ 
& = \Gamma \pi^{-1/2} \int_0^\infty dt \int_{\R^3} \frac{\mbox{\rm
    e}^{\,i\la \k,\,\x\ra - t |\x|^2 -t \mu^2}} {t^{1/2}}\,d\x \\ 
& = \Gamma \pi \int_0^\infty\frac{\mbox{\rm
    e}^{\,-|\k|^2/(4t) -t \mu^2}}{ t^{2}}\,dt \ge 0,
  \end{aligned}
\end{equation*}
where we used the fact that
$$ 
\lambda^{-1/2} = \pi^{-1/2} \int_0^\infty \text{e}^{\,-\lambda
  t}\,t^{-1/2} dt,
$$
and employed the explicit expression of the Fourier transform of the
Gaussian kernel.\par
Regarding (A.3) the integral over $\R^3$ can be split into the inner
and outer part. The first one is bounded since the function
$\widehat{\varphi}_R(\k)$ is $O(|\k|^{-2})$ near the origin, hence the
inner integral converges. On the other hand, the decay at infinity
necessary to show the convergence of the outer integral comes directly
by observing that $\widehat{\varphi}_R(\k)=2\mu^2\,\mbox{\rm
  Bessel}(1,2|\k|\mu) (\mu|\k|)^{-1}$. The fast decay at infinity
(enough to have both (A.3) and (A.4) satisfied) derives from the
properties of this special-function.\par     
In this case we obtain, by using as kernel the function $\varphi_R$,
exactly the Rosenhead model where the velocity
field is given by the equation:  
$$
\U(\x) = \Gamma \int_0^1 
\frac{\x-\bgamma(\xi)}{(|\x-\bgamma(\xi)|^2+\mu^2)^{3/2}} \wedge
\bgamma_\xi(\xi)\,d\xi.  
$$ 
\end{remark}
The evolution problem can be set up in different Banach spaces of
closed paths where local solution exists and are unique. For example
we can have solution for initial condition in the Sobolev space
$H^1_\#(0,1)$, in the space of H\"older continuous with exponent
greater than $1/2$, and also in some spaces of rough-paths,
see~\cite{bb,BGR}.\par  
Details of the proofs (and some specific assumptions) depend
however on the functional setting, so we will show the proofs in the
case of solutions living in $H^1_\#(0,1)$, using the framework of
Theorem~\ref{theorem_existence}. 
\section{Energy and global solutions}\label{section_energy}
For the evolution problem
\begin{equation}\label{eq:mod-evolution}
\frac{\partial}{\partial t} \bgamma(\xi,t) =\U(\bgamma(\xi,t),t).  
\end{equation}
 with
velocity given by~\eqref{eq:mod-velocity} we can identify a function
which plays the r\"ole of a ``kinetic'' energy.
\begin{definition}[Kinetic energy.]
The function below $\mathcal{H}:\bgamma\mapsto\R$, defined on the
space of smooth curves, is the ``kinetic energy'' for the smoothed
evolution problem associated to~\eqref{eq:mod-evolution}
\begin{equation}\label{eq:energy}
\mathcal{H}(\bgamma) := \frac{1}{2} \int_0^1 \int_0^1
\varphi(\bgamma(\xi)-\bgamma(\eta)) \la 
\bgamma_\xi(\xi),  \bgamma_\eta(\eta)\ra \,d\xi d\eta,   
\end{equation}
\end{definition}
The main result we shall prove is that $\mathcal{H}(\bgamma)$ is
constant in time.  Then, the existence of the energy allows to exploit
\textit{a-priori} estimates to have global existence for initial data
with finite energy. 

\begin{remark}
This particular expression for the energy has been
discussed also by Marsden and Weinstein~\cite{mw} and
Holm~\cite{Holm}, they show that on the space of closed curves in
$\R^{3}$ there exists a natural Poisson structure and that, formally,
with respect to this structure the function~(\ref{eq:energy}) is the
Hamiltonian function  which generates the flow described
by~(\ref{eq:mod-evolution}). 
\end{remark}
First, we show that the function $\mathcal{H}(\bgamma)$ is
well-defined.  
\begin{lemma}
For each smooth enough curve $\bgamma$, it holds that
$0\leq\mathcal{H}(\bgamma)<+\infty.$
\end{lemma}
\begin{proof}
In terms of Fourier variables, the energy $\mathcal{H}(\bgamma)$ can
be written as
\begin{equation*}
\begin{aligned}
\mathcal{H}(\bgamma)&=
\frac{1}{2(2\pi)^{3}}\int_0^1\int_0^1\int_{\R^3}
\widehat\varphi(\k)\,\text{e}^{-i\la\k,\,\bgamma(\xi)-
\bgamma(\eta)\ra}\, \bgamma_\xi(\xi)
\bgamma_\eta(\eta)\,d\k\,d\xi d\eta\\  
&=\frac{1}{2(2\pi)^{3}}\int_{\R^3} \widehat\varphi(\k) 
\left|\int_0^1\text{e}^{\,i\la \k,\,\bgamma(\xi)\ra}
\bgamma_\xi(\xi)\,d\xi\right|^2 d\k   
\end{aligned}
\end{equation*}
which, thanks to the assumption on the non-negativity of the Fourier
transform of $\varphi$, proves that the energy is a non-negative
quantity. 
Moreover, for any $\bgamma \in H^1_\#(0,1)$ we have the obvious
estimate 
$$
\left|\int_0^1 \text{e}^{\,i\la\k,\bgamma(\xi)\ra}\bgamma_\xi(\xi)
\,d\xi\right| \le \int_0^1 |\bgamma_\xi(\xi)|\,d\xi, 
$$ 
actually we observe that sharper estimates are possible, even if they
are not needed at this stage. Finally, we have proved that
$$ 
\mathcal{H}(\bgamma) \le
\frac{1}{2(2\pi)^{3}}\left(\int_{\R^3}\widehat\varphi(\k)\,d\k
  \right)\left(\int_0^1 |\bgamma_\xi(\xi)|\,d\xi\right)^2.
$$
The final observation is that both integrals are finite (use
assumption (A.3) and the fact that $\bgamma$ is rectifiable a curve),
showing that $\mathcal{H}(\bgamma)$ is a well-defined energy for any
$\bgamma \in H^1_\#(0,1)$.  
\end{proof}

\vspace{.2cm}

Next, we show that the function $\mathcal{H}(\bgamma)$ behaves as a
constant of motion if $\bgamma$ evolves under the flow associated to
$\U$.
\begin{lemma}
\label{lemma:const-energy}
Let $\bgamma(t,\xi)$ be a local smooth (as those of Th.~\ref{theorem_existence}) solution of the
problem~\eqref{eq:mod-evolution}, then
$$
\frac{d \mathcal{H}(\bgamma(t,\xi))}{dt}=0.
$$
\end{lemma}
\begin{proof}
We shall show by an explicit computation that the energy
$\mathcal{H}(\bgamma)$ is invariant and the completely anti-symmetric
Levi-Civita tensor $\epsilon_{ijk}$ will be used to write, with the
Einstein repeated indices convention, that
$$
\UU^i(\x) =\epsilon_{ijk}\int_0^1 \nabla^j \varphi(\x-\bgamma(\xi)) 
\bbgamma^k_\xi(\xi)\,d\xi. 
$$ 
Then, with explicit vector notations, we have:
\begin{equation*}
  \begin{aligned}
\frac{d \mathcal{H}(\bgamma(t,\xi))}{dt} & =  
\int_0^1 \int_0^1 \nabla^k \varphi({\bgamma(t,\xi)}-{\bgamma(t,\eta)})
 {\partial_t\bbgamma^k(t,\xi)}{\bbgamma_\xi^i(t,\xi)}
{\bbgamma_\eta^i(t,\eta)}\,d\xi d\eta\\  
& \qquad + \int_0^1 \int_0^1 \varphi({\bgamma(t,\xi)}-
{\bgamma(t,\eta)})
{\partial_t\bbgamma_\xi^i(t,\xi)}
{\bbgamma_\eta^i(t,\eta)} d\xi d\eta,
  \end{aligned}
\end{equation*}
where, for simplicity, $\partial_t
\bgamma(t):=\partial\bgamma(t)/\partial t$.\par  
By integrating by parts the $\xi$-integral in the second term we get
\begin{equation*}
  \begin{aligned}
\frac{d \mathcal{H}({\bgamma(t,\xi)})}{dt} & = \int_0^1 \int_0^1
 \nabla^k \varphi({\bgamma(t,\xi)}-{\bgamma(t,\eta)})
{\partial_t\bbgamma^k(t,\xi)}{\bbgamma_\xi^i(t,\xi)}
{\bbgamma_\eta^i(t,\eta)}\, d\xi d\eta \\ 
& \qquad - \int_0^1 \int_0^1 \nabla^k
 \varphi({\bgamma(t,\xi)}-{\bgamma(t,\eta)})
 {\partial_t\bbgamma^i(t,\xi)} {\bbgamma_\xi^k(t,\xi)}
 {\bbgamma_\eta^i(t,\eta)}\, d\xi d\eta \\ 
& = \epsilon_{cab} \epsilon_{cij} \int_0^1 \int_0^1 \nabla^a
 \varphi({\bgamma(t,\xi)}-{\bgamma(t,\eta)})
{\partial_t\bbgamma^i(t,\xi)}
{\bbgamma_\xi^j(t,\xi)}{\bgamma_\eta^b(t,\eta)}\, d\xi d\eta,   
  \end{aligned}
\end{equation*}
where we used the fact that 
$$
\epsilon_{cab}\epsilon_{cij}=\delta_{ai}\delta_{bj}-\delta_{aj}
\delta_{bi}. 
$$
Next, by definition of $\U$ we have
\begin{equation*}
  \begin{aligned}
\frac{d \mathcal{H}(\bgamma(t,\xi))}{dt} & = \epsilon_{cij} \int_0^1
\UU^c({\bgamma(t,\xi)}) {\partial_t\bbgamma^i(t,\xi)}
{\bbgamma_\xi^j(t,\xi)}\, d\xi\\ & = \epsilon_{cij} \int_0^1
\UU^c({\bgamma(t,\xi)})\UU^i({\bgamma(t,\xi)})
{\bbgamma_\xi^j(t,\xi)}\, d\xi\\ & = 0,
  \end{aligned}
\end{equation*}
where we used in sequence the equation~\eqref{eq:mod-evolution} of
motion for ${\bgamma(t)}$ and the complete anti-symmetry of the tensor
$\epsilon_{ijk}$.
\end{proof}

\vspace{.2cm}
The next step is to show that if the kinetic energy is bounded, then
the velocity $\U$ associated to the evolution problem is smooth. In
particular, it will follow that the velocity induced by vorticity
concentrated over a $H^1$-curve is very regular. The regularity of the
smoothing kernel is inherited by the velocity $\U$ even if the
framework is that of a singular problem. 
\begin{lemma}
\label{lemma:bound-u}
For any $0\le n \in\N$, we have the bound
\begin{equation}
\|\nabla^n \U\|_{L^\infty} \le \frac{1}{2\pi^{3/2}}\left[\int_{\R^3}
  |\k|^{2(1+n)} 
  \widehat\varphi(\k) \,d\k \right]^{1/2} \mathcal{H}^{1/2}(\bgamma), 
\end{equation}
provided that the integral $\int_{\R^3} |\k|^{2(1+n)}
  \widehat\varphi(\k) \,d\k$ is finite. 
\end{lemma}
\begin{proof}
The proof follows easily by using Cauchy-Schwartz inequality. In fact,
it follows that
\begin{equation*}
  \begin{aligned}
 |\U| & = \frac{1}{(2\pi)^3}\left|\int_{\R^3} \text{e}^{\,i\la \k,\x\ra}
 \widehat\varphi(\k)(i\k) \wedge \int_0^1
 \text{e}^{\,-i\la\k,\bgamma(\xi)\ra} \bgamma(\xi) \,d\xi d\k\right|\\
 & \le \frac{1}{(2\pi)^3}\left[\int_{\R^3} |\k|^2
 \widehat\varphi(\k)\,d\k\right]^{1/2} 
 \left[\int_{\R^3} \widehat\varphi(\k) \left|\int_0^1
 \text{e}^{\,-i\la \k,\bgamma(\xi)\ra} \bgamma(\xi)\,d\xi\right|^2
 d\k\right]^{1/2}\\ 
& = \frac{1}{2\pi^{3/2}}\left[\int_{\R^3} |\k|^2
 \widehat\varphi(\k)\,d\k \right]^{1/2} 
 \mathcal{H}^{1/2}(\bgamma),
  \end{aligned}
\end{equation*}
and in a similar fashion we can bound all the derivatives of $\U$,
provided that $\varphi$ is smooth enough: 
\begin{equation*}
\|\nabla^n \U\|_{L^\infty} \le \frac{1}{2\pi^{3/2}}\left[\int_{\R^3}
  |\k|^{2(1+n)} \widehat\varphi(\k) \,d\k \right]^{1/2}
\mathcal{H}^{1/2}(\bgamma). 
\end{equation*}  
\end{proof}

\vspace{.2cm}
Essentially we shall use this lemma just for $n=1$ to prove global
existence of solutions, hence assumption (A.4) will be enough. On the
other hand, to prove higher regularity of $\U$ these extra
conditions on $\widehat{\varphi}$ are needed.  
\vspace{.2cm}
%
%
\begin{theorem}
The evolution problem~\eqref{eq:mod-evolution} has a unique global
solution for any initial condition in $H^1_\#(0,1)$.
\end{theorem}
\begin{proof}
Denote $\vec{\boldsymbol\psi} \in H^1_\#(0,1)$ the initial condition,
then we already know that local solution $\bgamma(t,\xi)$ exists in a
(possibly small) time interval $t \in[0,T^*[$, depending on the
    $L^2$-norm of $\vec{\boldsymbol\psi}_\xi$, recall
Theorem~\ref{theorem_existence}. However, by taking the derivative
of~\eqref{eq:mod-evolution} with respect to $\xi$, and with and
integration over $[0,t]$, we have now the additional \textit{a-priori}
estimate ($\|\,.\,\|$ denotes the $L^2(0,1)$-norm) 
\begin{equation*}
  \begin{aligned}
\|\bgamma_\xi(t,\xi)\| & \le \|\bgamma_\xi(0)\| + \int_0^t \|\nabla
\U(s)\| \|\bgamma_\xi(s,\xi)\|\, ds \\ 
& \le \|\bgamma_\xi(0)\| + C\int_0^t \mathcal{H}^{1/2}(\bgamma(s))
\|\bgamma_\xi(s,\xi)\|\, ds \\ 
& = \|\bgamma_\xi(0)\| + C \mathcal{H}^{1/2}(\bgamma(0)) \int_0^t
\|\bgamma_\xi(s,\xi)\|\, ds,
  \end{aligned}
\end{equation*}
where the fist line is an easy bound for the norm of the solution, the
second line comes from Lemma~\ref{lemma:bound-u}, and the last line
from the fact that the energy is constant along solutions (as stated
in Lemma~\ref{lemma:const-energy}.)\par 
Then, by using the Gronwall inequality we have
$$ 
\sup_{t \in [0,T]}\|\bgamma_\xi(t,\xi)\| \le \|\bgamma_\xi(0)\|\,
\text{e}^{\,C \mathcal{H}^{1/2}(\bgamma(0))\,T},
$$
which guarantees existence of global solution, since any local
solution, having the $H^1(0,1)$ norm uniformly bounded can be uniquely 
continued up to any positive time $T$.
\end{proof}

\vspace{.2cm}
\subsection{H\"older initial conditions}
In~\cite{BGR} the authors prove that for sufficiently regular
functions $\varphi$ the vortex filament equation has local solutions
for initial conditions $\bgamma(0)$ which are H\"older continuous
functions of exponent $\alpha$ with $\alpha \ge 1/2$. In this case,
the line integrals of the form
\begin{equation}
\label{eq:line-integral}
\int_0^1 \langle\vec{\bf
   f}(\bgamma(t,\xi)),\,\bgamma_\xi(t,\xi)\rangle\, d\xi= 
   \int_0^1 \langle\vec{\bf f}(\bgamma(t,\xi)),\,d_\xi
  \bgamma(t,\xi)\rangle 
\end{equation}
must be understood as limits of Riemann sums: such limit exists if the
function $\vec{\bf f}: \R^3 \to \R^3$ is at least $C^1$ and this
process defines the integral in the sense of Young~\cite{Young}.\par  
Moreover, in the same paper, the authors extend the local existence
result, to a class of initial conditions living in a space of
\emph{rough paths} (see the book of Lyons for more
details~\cite{Lyons,LyonsBook}) that are special class of H\"older paths
with some additional structure. For rough paths
it is possible to define the line integrals (as those in
eq.~(\ref{eq:line-integral})) by means of natural ``renormalized''
Riemann sums. In this way it has been proved that the filament
equation (with sufficient regularity of the kernel $\varphi$) admits
local solutions starting from almost every 3D~Brownian loop
(i.e. curves chosen accordingly to the Wiener measure restricted to
the subspace of closed curves).\par 
The proof of the constancy of the energy extends also to these
different functional setting and the existence of global solutions can
be proved along the same strategy used for $H^1$ paths. We leave the
technical details to the interested reader, since they are outside the
scope of the present paper. For example, the extension of Lemma~\ref{lemma:const-energy} to H\"older solutions can be done using convergence of approximating discretizations and the computations contained in Sec.~\ref{discrete}.\par
\begin{remark}
The relevance of this additional result can be seen in the light of
the K41 theory and the Onsager conjecture on possible singularities in
the sense of $C^{0,\alpha}$ velocity fields. This may suggest that
velocity behaves as H\"older continuous velocity field (see~\cite{f}),
hence as a singularity diffused all over the fluid. More recent
developments link intermittency with concentration of singularities
on small sets, described by H\"older -hence (multi-)fractals- sets,
see~\cite{f,g}.
\end{remark}
\section{Impulse formulation}\label{Buttke_loops}
In this section we consider another singular distribution of vorticity
that is well-known and useful in the scientific computing, since it
involves vector fields (as the ``magnetization'') that are possibly
with non-vanishing divergence. In particular, we shall consider
\textit{Buttke loops} that (according to the terminology of
Chorin~\cite{MR95m:76043}) are small loops of vorticity in an
irrotational background, which evolve according to 3D~Euler
equation. The first study of the kinematic interaction of an
immersed body in an inviscid irrotational flow dates back to
Kelvin~\cite{Kelvin} and his analysis was based on the study of fluid
impulse. It was in that paper that he introduced a model of
``core-less vortices'' in order to explain some experimental facts on
eddy formation at the boundary.  The interaction of a vortex ring
(line) in an inviscid flow has been also studied by
Roberts~\cite{Roberts}, by using as canonical variables the position
of the centroid of each ring and its impulse (called by Roberts
``momentum of vorticity'').  A proper Hamiltonian formulation of the
3D~Euler equations have been 
discovered independently by different authors, see for instance
Osedelets~\cite{Osedelets} for a development in a continuum
setting, with canonical variables the position and the impulse
density. Then, around 1990 Buttke~\cite{MR1248421} linked the discrete
formulation of Roberts to fast and efficient discrete methods for the
numerical simulation of turbulent flows.  We consider also this
setting, since by using essentially the same techniques of the
previous section we are able to prove a global existence result also
for the latter model.\par
\subsection{Buttke loops and a discrete problem}
The main idea (refer for instance to \cite{MR1248421,MR95m:76043}) is
to introduce a new variable $\m$ (that is called \textit{magnetization} or
\textit{vortex magnetization}), that is obtained by adding to the 
velocity $\U$ a gradient at $t=0$:
\begin{equation}\label{impulse}
\m=\U+\nabla q.
\end{equation}
The unknown $\m$ does not satisfy the incompressibility constraint,
but it is with compact support and it satisfies
\begin{equation*}
\curl\U=\curl\m.
\end{equation*}
Then, we can interpret $\m$ as essentially local, while $\nabla q$
is an extensive field. The decomposition~\eqref{impulse} resembles the
Helmholtz decomposition, even if $\m$ is different from $\U$, or from
momentum density. The vector $\m$ is related to the so-called
``effective vorticity'' (compactly supported) by the relation
$\bxi=\curl\m$. Finally, the introduction of the new variable $\m$ can
be seen as a gauge transformation, known as ``geometric gauge.''\par  
The equation of evolution for $\m$ can be easily determined: 
\begin{equation*}
\frac{\partial\m}{\partial t}+(\U\cdot\nabla)\,\m+
\m\cdot(\nabla\U)^T={\bf0},
\end{equation*}
where $\U=\prob\m$, $\prob$ denoting the Leray projection over
divergence-free vector fields. The main point is that we have an
equivalent equation for $\U$ plus an arbitrary gradient at the initial
time. A proper choice of this gradient leads to the study of vortex
loops. In fact, if $\curl\U$ has support within a small ball $B$, then 
the resulting $\m$ has support within the same ball and 
\begin{equation*}
\m(\x)=\vec{\bf M}\,\rho(\x-{\x}_B)\quad\text{for some }{\x}_B\in B. 
\end{equation*}
Here $\vec{\bf M}$ is a vector in $\R^3$, while $\rho$ is a
non-negative smooth
function with support within $B$ and such that  
$\int_{\R^3}\rho(\x)\,d\x=1$. The ``magnet'' $\vec{\bf
  M}\,\rho(\x-{\x}_B)$ has a simple interpretation: the velocity field
induced is the same of a small vorticity loop (or vortex ring), with
$\vec{\bf M}$ perpendicular to the plane of the loop and $|\vec{\bf
  M}|=\Gamma \pi R^2$. Then, from the physical point of view the
vector $\m$ can be  interpreted as a vortex dipole density.\par  
The above interpretation can be used to deduce a Hamiltonian system
for a more complicated distribution of vorticity, see
Buttke~\cite{MR1248421}. This approach works as a de-singularization
of the problem: the resulting discrete dynamic we shall consider
includes in an implicit way a regularization that will allows us to
prove global existence of smooth solutions as well as good
computational properties investigated in~\cite{MR1248421}.\par
The discrete model is derived by considering a finite sum of magnets
and denoting by $\x_\alpha$, for $\alpha=1,\dots,N$, the positions of
the loops and by $\m_\alpha$ be the corresponding magnetization 
vector, that is not anymore a constant vector. Then, the velocity
field $\U$ of the fluid is given by the solution of the  
equation
\begin{equation*}
\U=\vec{\bf M}+\nabla q,
\end{equation*}
with
$$
\vec{\bf M}(\x,t)=\sum_{\alpha=1}^N \m_\alpha(t)\rho(\x-\x_\alpha),  
$$ 
where the scalar $\phi$ is determined by the equation
$$
\Delta q=-\nabla\cdot\vec{\bf M}.
$$
The equation of motion for $\vec{\bf M}$ is (component-wise)
\begin{equation*}
  \frac{\partial{\bf M}^i}{\partial t}+\sum_{j=1}^3\UU^j\,\nabla^j
  {\bf M}^i =-\sum_{j=1}^3{\bf M}^j\,\nabla^i \UU^j\qquad
  i=1,\dots,3
\end{equation*}
 and the ``loop particles'' moves according to the system of ordinary
differential equations
\begin{equation}
\label{eq:system1}
\left\{\begin{aligned}
&\dot {\bf m}_\alpha^i = - \sum_{j=1}^3{\bf m}_\alpha^j \nabla^i
  \UU^j(\x_\alpha)\\ 
&\dot {\bf x}_\alpha^i = \UU^i(\x_\alpha),
       \end{aligned}
\right.
\end{equation}
where the index $\alpha$ run over particle labels.\par
Equations~\eqref{eq:system1} form a Hamiltonian system, if we consider
the conjugate pairs of variables
$(\x_\alpha,\m_\alpha)_{\alpha=1,\dots,N}$ with symplectic structure 
$$
d\Omega = \sum_{\alpha=1}^N\sum_{i=1}^3 d{\bf m}_\alpha^i \wedge d{\bf
  x}_\alpha^i 
$$ 
and Hamiltonian function
\begin{equation*}
\mathcal{H}(\{\x_\alpha,\m_\alpha\})=\frac{1}{2} 
\sum_{\alpha,\beta=1}^N \m_\alpha \cdot \m_\beta\
\rho(\x_\alpha-\x_\beta) + (\m_\alpha \cdot \nabla) (\m_\beta\cdot
\nabla) \Phi(\x_\alpha-\x_\beta),   
\end{equation*}
where $\Phi$ is the solution of $\Delta\Phi=\rho$. Hence
system~\eqref{eq:system1} takes the canonical form  
\begin{equation*}
\left\{\begin{aligned}
&{\dot{\x}}_\alpha=\frac{\partial \mathcal{H}}{\partial\m_\alpha}
\\
&{\dot{\m}}_\alpha=-\frac{\partial \mathcal{H}}{\partial {\x}_\alpha}.
       \end{aligned}
\right.
\end{equation*}
\begin{remark}
The analogy between magneto-static and fluid mechanics is that to the
electric current it corresponds the vorticity and to the magnetic
induction it corresponds the velocity. Since the name
``magnetization'' may be misleading, we prefer to use Chorin's
notation and to refer to them as Buttke loops or vorticity loops.  
\end{remark}
In order to show that the energy is constant on solutions we pass to
the wave-numbers notation. Let $\widehat\rho(\k)$ be the Fourier
transform of $\rho(\x)$. The energy
$\mathcal{H}(\{\x_\alpha,\m_\alpha\})$ can be written, in terms of
wave-numbers, as follows 
\begin{equation*}
\mathcal{H}(\{\x_\alpha,\m_\alpha\}) = \frac{1}{2(2\pi)^3}\int_{\R^3}
\widehat{\rho}\,(\k) 
\left|\Pi_{\k}\, \sum_{\alpha=1}^N \m_\alpha
\,\text{e}^{\,i\la\k\,\x_\alpha\ra} \right|^2\,d\k,
\end{equation*}
where $\Pi_{\k}$ is the projection in the plane orthogonal to the
vector $\k$. From this representation it is clear that the energy is
positive definite. Moreover, the velocity field has the form
\begin{equation*}
\U(\x) = \frac{1}{(2\pi)^3}\int_{\R^3} \widehat{\rho}\,(\k)\
\Pi_{\k}\ \sum_{\alpha=1}^N  
\m_\alpha\, \text{e}^{\,-i\la\k,\,\x-\x_\alpha\ra}\, d\k
\end{equation*}
and the tensor $\nabla\U$ appearing in the equation for $\m$ is
\begin{equation*}
[\nabla \U]^{lj}(\x)=\frac{1}{(2\pi)^3}\int_{\R^3}
\widehat{\rho}\,(\k) (-i\k)^l   
\sum_{\alpha=1}^N (\Pi_{\k}\ \m_\alpha)^j
\ \text{e}^{\,-i\la\k,\,\x-\x_\alpha\ra}\, d\k.    
\end{equation*}

\subsection{The periodic problem}
In order to avoid problems related with the decay at infinity and
possibly non-converging integrals, we restrict to the periodic
setting. Thus, we fix a box of linear size $2\pi$ and consider the
periodic version of the above problem, i.e., we substitute Fourier
series expansion to the Fourier transform.\par
The energy is now given by
\begin{equation*}
  \mathcal{H}(\{\x_\alpha,\m_\alpha\}) = \frac{1}{2(2\pi)^3}\sum_{\k\in\Z^3}
  \,\widehat{\rho}\,(\k)\, 
  \left|\Pi_{\k}\ \sum_{\alpha=1}^N \m_\alpha \,\text{e}^{\,i\la \k,\,
  \x_\alpha\ra} \right|^2, 
\end{equation*}
where $\k$ runs over $\Z^3$ and the coordinate $\x_\alpha$ is
restricted to the box $\Lambda_L = ]-\pi,\pi[^3$. The velocity is then 
\begin{equation*}
\U(\x)=\frac{1}{(2\pi)^3}\sum_{\k\in\Z^3}\widehat{\rho}\,(\k)\
\Pi_{\k}\ \sum_{\alpha=1}^N \m_\alpha \,\text{e}^{\,-i\la
  \k,\,\x-\x_\alpha\ra}\qquad \forall\,\x \in ]-\pi,\pi[^3.    
\end{equation*}
By using standard techniques we are able to prove the following
result.
\begin{proposition}\label{bump}
The solution for the evolution problem of a finite number of loops in
the periodic box $\Lambda_L$ exists (and is unique) for any positive
time, provided that the initial condition has finite energy and the
function $\widehat\rho(\k)$ is non-negative and satisfies the decay
estimate  
$$
\sum_{\k\in\Z^3} |\k|^4\, \widehat{\rho}\,(\k) < +\infty.
$$   
\end{proposition}
\begin{remark}
The condition of decay for $|\k|\to+\infty$ required in
Proposition~\ref{bump} is easily satisfied if $\rho(\x)$ is a smooth
enough function. The condition on the non-negativity of
$\widehat{\rho}\,(\k)$ is not 
automatically satisfied for each smooth function $\rho(\x)$. A wide
class of smooth functions with compact support and non-negative (hence
real) Fourier transform can be identified as follows. Consider a
smooth (say $C^\infty$) ``bump'' function $\textsl{b}(\x)$ over the
ball $B({\bf0},\delta)$ and, in addition, suppose that
$\textsl{b}(\x)$ is ``even,'' implying that the Fourier transform has
vanishing imaginary part. Then, define
$\rho(\x):=(\textsl{b}*\textsl{b})(\x)$ and this 
will turn out to be smooth, non-negative, and null outside the ball
$B({\bf0},2\delta)$ of radius twice that of the original one. Finally,
by using the convolution theorem it will follow that 
$$
\widehat{\rho}\,(\k)=
\widehat{\textsl{b}\ast\textsl{b}}\,(\k)=
\widehat{\textsl{b}}(\k)\cdot\widehat{\textsl{b}}(\k)\geq 0.  
$$
\end{remark}
\begin{proof}[Proof of Proposition~\ref{bump}] 
Again the proof is based on an energy estimate and an
\textit{a-priori} bound of the velocity in terms of the energy. The
main point of the proof is the following Cauchy-Schwartz inequality:
\begin{equation*}
  \begin{split}
|\U(\x)| & \le \frac{1}{(2\pi)^3}\sum_{\k\in\Z^3}
\widehat{\rho}\,(\k)\left|\Pi_{\k}\  
\sum_{\alpha=1}^N \m_\alpha \,\text{e}^{\,-i\la
  \k,\,\x-\x_\alpha\ra}\right|\\  
& \le \frac{1}{(2\pi)^3}\left[\sum_{\k\in\Z^3}
  \widehat{\rho}\,(\k)\right]^{1/2} 
\left[\sum_{\k\in \Z^3} \widehat{\rho}\,(\k) \left|\Pi_{\k}
\sum_{\alpha=1}^N \m_\alpha\,\text{e}^{\,-i\la
  \k,\,\x-\x_\alpha\ra}\right|^2 \right]^{1/2} \\  
& = \frac{1}{2\pi^{3/2}}\left[\sum_{\k\in\Z^3}
  \widehat{\rho}\,(\k)\right]^{1/2} 
\mathcal{H}^{1/2}(\{\x_\alpha,\m_\alpha\}).
  \end{split}
\end{equation*}
and in the same way it easily follows that
\begin{equation*}
  \begin{split}
 \|\U\|+\|\nabla \U\|+\|\nabla^2
   \U\|\le\frac{1}{2\pi^{3/2}}\left[\sum_{\k\in \Z^3} |\k|^4 
   \widehat{\rho}\,(\k)\right]^{1/2}
   \mathcal{H}^{1/2}(\{\x_\alpha,\m_\alpha\}), 
  \end{split}
\end{equation*}
showing that we can control the regularity of $\U$ in terms of
the energy $\mathcal{H}(\{\x_\alpha,\m_\alpha\})$. By a direct
differentiation it is also clear that
$\mathcal{H}(\{\x_\alpha,\m_\alpha\})$ is constant along (smooth
enough) solutions of the equations~\eqref{eq:system1} so that we have
good control over $\U$ for any time. This implies the global existence
of solutions.
\end{proof}

\vspace{.2cm}
\section{A class of discrete models of evolution of vorticity}
\label{discrete} 
In this section we study a class of discrete models for the evolution
of vorticity which includes discrete versions of the line vortex model.
These models could be of interest in the numerical simulation using
lattice methods as suggested in~\cite{MR95m:76043}. In the sequel we
shall always assume on the function $\varphi$ the four natural
hypotheses (A.1)-(A.4), we introduced in
Section~\ref{regularized_problems}.\par 
%
We consider distribution of vorticity giving rise to a velocity
field of the form
\begin{equation*}
\U(\x) = \sum_{\alpha=1}^N \nabla \varphi(\x - \x_\alpha)\,\wedge
\bxi_\alpha 
\end{equation*}
that is parametrized by the set of $6N$ variables
$\{\x_\alpha,\bxi_\alpha\}_{\alpha=1,\dots,N}$. The physical
interpretation of this equation is that of describing the
superposition of $N$ vortex ``blobs'' (i.e. concentrations of
vorticity) indexed by $\alpha$, situated at points $\x_\alpha$, and
whose vorticity is directed along the vector $\bxi_\alpha$. In this
case these blobs determine the initial datum and there are no
assumptions that the solution persists being of the same type for
all positive times. This is the difference between blobs and loops
considered in the previous section.\par
%
%
\textit{A priori} we can think to each of the variables $\{\x_\alpha,
\bxi_\alpha\}_\alpha$ as an independent degree of freedom. However, if
we set $\bxi_\alpha = \x_{\alpha+1} - \x_{\alpha}$ (with the
understanding that $\alpha +1 = 0$ if $\alpha = N$), then this model
corresponds to a natural discretization of the line vortex model
previously described.\par
For the moment we do not impose any constraint on the variables $\bxi$
(either static or dynamic) and we just impose that the points
$\x_\alpha$ are transported by the flow $\U$ so that we have
\begin{equation}
  \label{eq:dyn-of-points}
{\dot{\x}}_\alpha = \U(\x_\alpha), \qquad \alpha=1,\dots,N.
\end{equation}
This is the discrete counterpart of~\eqref{characteristic}. %
If we want to address the problem of the global (in time) dynamics  of
these models we can introduce the following useful quantities
\begin{equation*}
\mathcal{L} = \sum_{\alpha=1}^N |\bxi_\alpha|\qquad\text{and}\qquad
\mathcal{A} = \sum_{\alpha=1}^N |\bxi_\alpha|^2  
\end{equation*}
that are respectively the length and the quadratic variation of the
curve itself.  Moreover, in analogy with the Hamiltonian introduced in
Section~\ref{section_energy}, we define an energy function (which
controls the magnitude and regularity of the velocity field $\U$) as
follows:
\begin{equation*}
\mathcal{H}(\{\x_\alpha, \bxi_\alpha\}) = \frac{1}{2}\sum_{\alpha,\beta=1}^N
\varphi(\x_\alpha-\x_\beta) \,\la\bxi_\alpha,\,\bxi_\beta\ra. 
\end{equation*}

It is not difficult to show that we have an analogous of
Lemma~\ref{lemma:bound-u}, which allows us to control $\U$ in term of
$\mathcal{H}$ and, of course, we have the following trivial bounds
\begin{equation*}
\|\nabla^n \U\|_{L^\infty} \le c_{n,\varphi} \mathcal{L}, \quad
\text{and}\quad
\mathcal{H} \le \|\varphi\|_{L^\infty} \mathcal{L}^2,  
\end{equation*}
where the constants $c_{n,\varphi}$ depends only on $\varphi$ and the
number $n$ of derivative (we assume that $\varphi$ is
sufficiently regular for all the constants in this section to be well
defined. For example it is enough to take $\varphi$ infinitely
differentiable.)\par
Let us now compute the time derivative of the energy. Taking into
account eq.~\eqref{eq:dyn-of-points}, but leaving the chance of an
arbitrary time-derivative $\partial_t\bxi_\alpha$ for $\bxi_\alpha$,
we obtain 
\begin{equation*}
  \begin{split}
 \frac{d\mathcal{H}(\{\x_\alpha, \bxi_\alpha\})}{dt}&
=\sum_{\alpha,\beta=1}^N 
\varphi(\x_\alpha-\x_\beta)\la\bxi_\alpha,\,\partial_t\bxi_\beta\ra-
\sum_{\alpha,\beta=1}^N \U(\x_\beta)\cdot \nabla 
\varphi(\x_\alpha-\x_\beta) \la\bxi_\alpha,\,\bxi_\beta\ra.
\end{split} 
\end{equation*}
We now analyze the second sum appearing in the above equality, in
order to recover some symmetries or cancellation properties. We get
\begin{equation*}
  \begin{split}
I & = \sum_{\alpha,\beta=1}^N \U(\x_\beta)\cdot \nabla
 \varphi(\x_\alpha-\x_\beta)
 \la\bxi_\alpha,\,\bxi_\beta\ra-\sum_{\alpha,\beta=1}^N
 \bxi_{\alpha}\cdot \U(\x_\beta)\, \bxi_\beta\cdot \nabla
 \varphi(\x_\alpha-\x_\beta) \\ 
& = \sum_{\alpha, \beta=1}^N \langle\bxi_\alpha \wedge \nabla
\varphi(\x_\alpha-\x_\beta),\,\bxi_\beta\wedge\U(\x_\beta)\rangle\\ 
&=-\sum_{\beta=1}^N
 \la\U(\x_\beta),\,[\bxi_\beta\wedge\U(\x_\beta)]\ra=0, 
  \end{split} 
\end{equation*}
and this implies that we can write the following equivalent expression
for the time derivative of $\mathcal{H}(\{\x_\alpha,\bxi_\alpha\})$: 
\begin{equation*}
  \begin{split}
 \frac{d\mathcal{H}(\{\x_\alpha,\bxi_\alpha\}) }{dt}& =
\sum_{\alpha,\beta=1}^N \varphi(\x_\alpha-\x_\beta)
\la\bxi_\alpha,\,\partial_t\bxi_\beta\ra -\sum_{\alpha,\beta=1}^N
\la\bxi_{\alpha},\,\U(\x_\beta)\ra\ \la\bxi_\beta ,\,\nabla
\varphi(\x_\alpha-\x_\beta)\ra.  
  \end{split}
\end{equation*}
Consider the second term and write
\begin{equation*}
  \begin{split}
A & = \sum_{\alpha,\beta=1}^N \la\bxi_{\alpha},\,\U(\x_\beta)\ra\ 
\la \bxi_\beta,\,\nabla \varphi(\x_\alpha-\x_\beta)\ra \\ 
& = - \sum_{\beta=1}^N \la\bxi_\beta,\,\nabla \Phi(\x_\beta)\ra +
 \sum_{\alpha,\beta=1}^N \sum_{i,j=1}^3 \varphi(\x_\alpha-\x_\beta)\,
 \bbxi_{\alpha}^i \nabla^j\UU^i(\x_\beta)\,\bbxi_\beta^j,
  \end{split}
\end{equation*}
where
$$ 
\Phi(\x) = \sum_{\alpha=1}^N \varphi(\x-\x_\alpha)
\la\bxi_{\alpha},\,\U(\x)\ra. 
$$
Finally, the rate of change of the energy takes the form
\begin{equation*}
\begin{split}
\frac{d}{dt}\mathcal{H}(\{\x_\alpha, \bxi_\alpha\}) & =
\sum_{\alpha,\beta=1}^N\sum_{i,j=1}^3  \varphi(\x_\alpha-\x_\beta)\, \bbxi_\alpha^i
\left[ {\partial_t \bbxi}_\beta^i -\nabla^j \UU^i(\x_\beta)
  \,\bbxi_\beta^j\right] + \sum_{\beta=1}^N  \bxi_\beta \cdot\nabla
\Phi(\x_\beta).  
  \end{split}
\end{equation*}
We can rewrite the above expression by setting
$\vPsi(\x) = \sum_{\alpha=1}^N \varphi(\x-\x_\alpha)\, \bxi_{\alpha}$
hence obtaining
\begin{equation}
\label{eq:energy-rate}
  \begin{split}
 \frac{d}{dt}\mathcal{H}(\{\x_\alpha , \bxi_\alpha\}) & =
 \sum_{\beta=1}^3\sum_{i,j=1}^3
\boldsymbol{\Psi}^i(\x_\beta) \left[ {\partial_t \bbxi}_\beta^i -
\nabla^j \UU^i(\x_\beta)\,\bbxi_\beta^j\right]+ 
\sum_{\beta=1}^N \bxi_\beta \cdot \nabla\Phi(\x_\beta), 
  \end{split}
\end{equation}
with $\Phi = \vPsi \cdot \text{curl\,} \vPsi$ since, by definition,
$\U(\x) = \text{curl\,} \vPsi(\x)$. By using a first order Taylor
expansion with integral remainder, the introduction of the auxiliary
function $\Phi$ allows us to rewrite the second term
of~\eqref{eq:energy-rate} as follows:  
\begin{equation}
\label{eq:int-by-parts}
\sum_{\beta=1}^N \bxi_\beta \cdot \nabla\Phi(\x_\beta) = 
  \sum_{\beta=1}^N [\Phi(\x_\beta+\bxi_\beta) - \Phi(\x_\beta)]
+ \sum_{\beta=1}^N \la \bxi_\beta, \Theta_\beta \bxi_\beta\ra,
\end{equation}
where $\Theta_\beta$ is the matrix
\begin{equation*}
\Theta_\beta^{ij} = \int_0^1 (1-\rho) 
\nabla^j \nabla^i \Phi(\x_\beta+\rho\,
\bxi_\beta)\,d\rho.
\end{equation*}
We observe that analogously to Lemma~\ref{lemma:bound-u} we
can control the function $\vPsi$ in terms of the energy $\mathcal{H}$
according to the following lemma, whose straightforward proof is left
to the reader.  
\begin{lemma}
\label{lemma:bound-psi}  
For any $0\le n \in\N$, we have the bound
\begin{equation*}
\|\nabla^n \vPsi\|_{L^\infty} \le \left[\int_{\R^3} |\k|^{2n}
  \widehat\varphi(\k) \,d\k \right]^{1/2} \mathcal{H}^{1/2}
(\{\x_\alpha,\bxi_\alpha\}).  
\end{equation*}
\end{lemma}

\vspace{.3cm}
\noindent In the same way we can bound $\Phi$ since, e.g., it holds
\begin{equation*}
 \|\Phi\|_{L^\infty} =  \|\vPsi \cdot \curl \vPsi\|_{L^\infty} 
\le c_\varphi \mathcal{H},    
\end{equation*}
where $c_\varphi$ is some positive constant depending just on the
Sobolev regularity  of $\varphi$.\par   
At this point (see also~\cite{MR95m:76043}) we may recognize to have
two interesting cases: 
\begin{itemize}
\item[A)] The \emph{discrete filament model}, obtained by setting
  $\bxi_\alpha = \x_{\alpha+1}-\x_\alpha$. In this case the first term
  in the r.h.s. of  Eq.~\eqref{eq:int-by-parts} vanishes due to the
  fact that
$$
\sum_{\beta=1}^N [\Phi(\x_\beta+\bxi_\beta) -
  \Phi(\x_\beta)]=\sum_{\beta=1}^N [\Phi(\x_{\beta+1}) -
  \Phi(\x_\beta)] = 0. 
$$  
\item[B)] A \emph{blob model}, obtained by considering each couple
  $(\x_\alpha, \bxi_\alpha)$ as describing (an approximation of) a 
vortex blob and for which the dynamics of the $\bxi_\alpha$ is fixed
by prescribing that these vectors are transported by the flow at the
point $\x_\alpha$, i.e.
\begin{equation*}
\dot{\bxi}_\alpha = (\nabla \U(\x_\alpha))^T \cdot\bxi_\alpha.
\end{equation*}
In this case the first term in the r.h.s. of
Eq.~\eqref{eq:energy-rate} vanishes identically.  
\end{itemize}
For these two models we prove the following result:
\begin{lemma}
\label{lemma:H-bounds}
In the discrete filament model we have
\begin{equation*}
  |\partial_t \mathcal{H}| \le c_1 \mathcal{H} \mathcal{A} + c_2
  \mathcal{H}^{1/2} \mathcal{A}. 
\end{equation*}
In the blob model we have
\begin{equation}\label{eq:energy-bound-blob}
  |\partial_t \mathcal{H}| \le  c_3 N \mathcal{H}^{1/2} + c_4
  \mathcal{H}^{1/2} \mathcal{A}
\qquad
  |\partial_t \mathcal{H}| \le  c_5 \mathcal{L} \mathcal{H}^{1/2}. 
\end{equation}

Moreover, in both models we have
\begin{equation*}
 |\partial_t \mathcal{A}| \le c_6 \mathcal{H}^{1/2}  \mathcal{A} 
\qquad
 |\partial_t \mathcal{L}| \le c_7 \mathcal{H}^{1/2}  \mathcal{L}. 
\end{equation*}
The positive constants $c_i$, for $i=1,\dots,7$, depend only on the
function $\varphi$.
\end{lemma}
\begin{proof}
Given Lemma~\ref{lemma:bound-psi}, the only nontrivial part is to
justify the first term appearing in the bound for the discrete
filament model, i.e., $c_1 \mathcal{H} \mathcal{A}$. This term comes
form the expression  
$$
G =  \sum_{\beta=1}^N\sum_{i,j=1}^3\boldsymbol{\Psi}^i(\x_\beta)
\left[ {\partial_t \bbxi}_\beta^i - 
\nabla^j \UU^i(\x_\beta)\,\bbxi_\beta^j\right]
$$
appearing as the first term in the r.h.s. of
eq.~\eqref{eq:energy-rate}. In the discrete filament model we have
(note that $\bxi$ depends just on $t$, hence the partial derivative
is in fact a total derivative) 
\begin{equation*}
\dot{\bxi}_\alpha =
{\dot{\x}}_{\alpha+1}-{\dot{\x}}_\alpha=\U(\x_{\alpha+1})-\U(\x_\alpha)   
\end{equation*}
and, by a first order Taylor expansion, we get
\begin{equation*}
\dot{\bxi}_\alpha =   [\nabla \U(\x_\alpha)]^T \cdot \bxi_\alpha +
 \widetilde \Theta_\alpha (\bxi_\alpha \otimes
\bxi_\alpha), 
\end{equation*}
where the tensor $\widetilde \Theta_\alpha$ is given by the formula
\begin{equation*}
 \widetilde \Theta_\alpha^{kij} = \int_0^1 (1-\rho) \,
\nabla^i \nabla^j \U^k(\x_\alpha+\rho\, \bxi_\alpha)\,d\rho . 
\end{equation*}

Then, we can bound $G$ as follows:
\begin{equation*}
  \begin{split}
|G| & \le \|\vPsi\|_{L^\infty}\sum_{\beta=1}^N \sum_{i,j=1}^3| {\partial_t
  \bbxi}_\beta^i-\nabla^j \UU^i(\x_\beta)\,\bbxi_\beta^j | 
= \|\vPsi\|_{L^\infty}  \sum_{\beta=1}^N |
 \widetilde \Theta_\beta (\bxi_\beta \otimes
\bxi_\beta) 
|
\\ & \le  \|\vPsi\|_{L^\infty}  \sum_{\beta=1}^N |\bxi_\beta|^2
 \sup_{ijk}|\widetilde \Theta_\beta^{ijk}| 
\le  \|\vPsi\|_{L^\infty}  \|\nabla \U\|_{L^\infty} \sum_{\beta=1}^N
|\bxi_\beta|^2  
\\ & \le c_1 \mathcal{H} \mathcal{A}.
  \end{split}
\end{equation*}
The other bounds  are proved similarly.
\end{proof}

\vspace{.2cm}
\begin{remark}
We note that the term $c_3 N \mathcal{H}^{1/2}$ in the rate of energy 
change of the blobs model derives from the trivial bound
$$
\left|\sum_{\beta=1}^N
	  [\Phi(\x_{\beta}+\bxi_\beta)-\Phi(\x_\beta)]\right| \le 
2 N \|\Phi\|_{L^\infty}
$$ 
which, however, misses any possible cancellation property of the various
terms. Heuristically we expect that, if the evolution is constrained in
a bounded volume (e.g. by imposing periodic boundary conditions on a
torus), the distribution of the set of points $\{\x_\alpha\}_\alpha$
should not differ significantly from that of the points
$\{\x_\alpha+\bxi_\alpha\}_\alpha$ and thus that the two sums
$\sum_\alpha \Phi(\x_\alpha)$ and $\sum_\alpha
\Phi(\x_\alpha+\bxi_\alpha)$ should not be too different (at least not
of the order of $N$). 
\end{remark}

\subsection{Heuristics about adaptive numerical algorithms for vortex filaments}
The results of the previous section suggest a strategy to implement a
numerical algorithm which approximates the global solution of the
filament equation. Here we do not pretend mathematical rigor but only lay down some possible developments of the main results of this paper.

So let us imagine to start a numerical approximation of the vortex model with some arbitrary discretization level (with
$N$ points) and  follow the solution up to the instant when the variable
$\mathcal{A}(t)$ is larger than a fixed value, e.g. $1$. At this point
 proceed to a refinement of the discretization up to level $M>N$ in
order to ensure that $\mathcal{A}(t)$ become lower than $1/2$ and 
continue the simulation. Upon increased discretization the energy change, but it is not difficult to argue that this change will be small (since in the limit of infinite discretization the energy will converge to a limit). So in first approximation we will neglect this change (actually is possible to devise a discrete model where the filament is replaced by a piece-wise approximation, in which, upon refinement, the energy does not change at all). In this way we ensure that, at any
time $\mathcal{A}(t) \le 1$ and so we have also an (approximate) bound for the
growth of $\mathcal{H}$:
$$
\mathcal{H}(t) \le \mathcal{H}(0) \,\text{e}^{\,\max\{c_1,c_2\} t} 
$$ 
which implies that there will be no blow-up of the solution. Of course
since $|\partial_t \mathcal{L}| \le c\, \mathcal{H}^{1/2} \mathcal{L}$
there cannot be blow-up for the length of the discrete filament as
long as the energy remains finite.\par
This algorithm is very similar to the heuristic techniques that are
already adopted in numerical simulations of vortex filament dynamics,
but here we can identify a clear quantitative criterion when to perform the
refinement.\par
Recall also that the rate of change of energy is, in this case, given by
\begin{equation*}
    \begin{split}
 \frac{d}{dt}\mathcal{H}(\{\x_\alpha , \bxi_\alpha\}) & =
 \sum_{\beta=0}^N \la \vPsi(\x_\beta),  \widetilde \Theta_\beta
 \bxi_\beta\otimes\bxi_\beta \ra+ \sum_{\beta=0}^N \la \bxi_\beta,
 \Theta_\beta \bxi_\beta\ra, 
  \end{split}
\end{equation*}
cfr. eq.~(\ref{eq:energy-rate}) and Lemma~\ref{lemma:H-bounds}.  Then,
a more refined algorithm to decide where to increase the resolution of
the discretization is to find points along the filament where the 
quantity
$$
J_\beta = 
\la {\vPsi}(\x_\beta),  \widetilde \Theta_\beta
\bxi_\beta\otimes\bxi_\beta \ra
+  \la \bxi_\beta, \Theta_\beta \bxi_\beta\ra
$$ 
is large. These points are responsible for large positive
contributions to the change of energy and thus can be associated to
points of instability of the finite-element approximation of the
dynamics.  By refining the mesh only in the neighborhood of these
points we diminish the rate of change of the energy and control the
discrete evolution. Currently we do not know if the latter refinement
technique is too expensive from the computational point of view. \par
Finally, we observe that the presence of $N$ in the first term on the
r.h.s of~\eqref{eq:energy-bound-blob} prevents to obtain the same
result for the vortex blob model, since in this case refinement will
enlarge this term.
\section{Conclusions}
We considered several classes of problems related to the
evolution of special patterns of singular vorticity. Various kind
of regularization have been analyzed and they are based either 1) on a
smoother version of the Biot-Savart law, linking velocity and
vorticity or either 2) a ``point vortex'' special expression of the  
solution. Under reasonable assumptions we have been able to prove
global-in-time existence of solutions for these various models,
employing the Hamiltonian structure of the underlying 3D~Euler
equations. \par 
In the case of the vortex filament we have also been able to consider
curves that are not very smooth, solving an approximation of the line
vortex equation containing also the classical Rosenhead approximation
for this problem. In the case of small vortex loops we solved a periodic
problem involving a finite number of them.\par 
In the final section we used these results to obtain, as by-product,
some results on a couple of discrete models. The discrete models we
have been able to analyze are a discrete filament model and a model
of 3d vorticity blobs. For the latter we have been unable to find
global existence results, while for the former we linked global
regularity of the limit problem (smooth curve) with a numerical
procedure to properly simulate this vorticity pattern. In 
particular, a quantitative refinement technique has been suggested, in
order to produce an effective numerical simulation for arbitrarily
large times. 
\providecommand{\bysame}{\levaevmode\hbox to3em{\hrulefill}\thinspace}
\providecommand{\MR}{\relax\ifhmode\unskip\space\fi MR }
\providecommand{\MRhref}[2]{%
  \href{http://www.ams.org/mathscinet-getitem?mr=#1}{#2}
}
\providecommand{\href}[2]{#2}

\end{document}